\documentstyle[12pt]{article}
\author{Hans - J\"urgen Schmidt}
\title{A new proof of
 Birkhoff's theorem\footnote{extended version of a lecture 
read at the university of Cagliari/Italy April 22, 1997}
}
\date{}
\begin{document}
\maketitle

\centerline{
Universit\"at  Potsdam, Institut f\"ur Mathematik}
\centerline{
      D-14415 POTSDAM, PF 601553, Am Neuen Palais 10, Germany}

\begin{abstract}

Assuming $SO(3)$-spherical symmetry, the 4--dimensional 
Einstein equation reduces to an equation conformally
related to the field equation for 2--dimensional gravity 
following  from  the Lagrangian $L  =  \vert R \vert^{1/3}$.

Solutions for 2--dimensional gravity always possess
a local isometry because the traceless part of its Ricci tensor
identically vanishes. 

Combining both facts, we get a new proof of Birkhoff's theorem;
contrary to other proofs, no coordinates must be introduced.

The  $SO(m)$-spherically symmetric solutions of the
 $(m+1)$--dimen\-sio\-nal Einstein equation can be found by
considering $L \, = \, \vert R \vert^{1/m}$ in
two dimensions. This yields several generalizations 
of Birkhoff's theorem in an arbitrary number of dimensions,
and to an arbitrary signature of the metric. 

\end{abstract}

\medskip
 
 to appear in {\it Grav. and Cosmol.}

\bigskip

\section{Introduction}
\setcounter{equation}{0}

The Birkhoff theorem states that every spherically symmetric 
vacuum solution of Einstein's general relativity theory 
 is part of the Schwarzschild solution; therefore, the solution  
 possesses a four-dimensional\footnote{This is the remarkable
part of that theorem, because the rotation group 
$SO(3)$ is only 3--dimensional.}
isometry group, at least locally. 
This theorem has been generalized into several directions,
 e.g. to  Einstein--Maxwell fields and to 
higher dimensions  by Bronnikov,
Kovalchuk and Melnikov in [1].

In the meanwhile it is generally known that the popular  
formulation ``every spherically symmetric 
vacuum solution is static'' is misleading, because
inside the horizon, the Schwarzschild  black hole is not static.
Now, Bondi and Rindler [2] argued, that even in the absence of
a horizon, the use of the word ``static'' might be
misleading, too. 
 
Both Schmutzer and Goenner [3] cite refs. [4] for the 
original Birkhoff theorem, the historic development 
can be summarized as follows: ``Jebsen (1921)
 was the first to formulate it and Birkhoff (1923) 
was the first to prove it''.  
 Hawking, Ellis [5, page 369] gave
a proof of it based on B. Schmidt's method [6]. A
further proof is given in sct. 32.2 of Misner, Thorne and
Wheeler [7]. They relate this 
theorem to the fact that no monopole gravitational waves exist in
Einstein's theory. 
The Birkhoff theorem in 2--dimensional pure metric gravity has
been deduced in [8] and generalized in [9]. It possesses
analogous formulations for 2-dimensional dilaton--gravity
theories, cf. refs. [10 - 14] for its recent development.

In ref. [15], Ashtekar, Bicak and B. Schmidt considered 
those solutions of the 4--dimensional Einstein equation which
possess one translational Killing field. The non-trivial
solutions fail to be asymptotically flat; to get nevertheless 
a sensible notion of energy it turned out  to be 
advantageous to reformulate the problem as 3--dimensional
gravity with additional matter.

At the same formal level, but with another scope as in [15],
Romero, Tavakol and Zalaletdinov [16] reduced the dimension from 
5 to 4. One of their results read: 4--dimensional gravity with
matter can be isometrically embedded into a 5--dimensional 
Ricci--flat space--time, i.e., into a vacuum solution of the 
 5--dimensional Einstein equation. 

\medskip

In the present paper, we
 follow a similar line as in [15] and [16] and  
 deduce  Birkhoff--type theorems for
warped products of manifolds where one of them is 
two--dimensional. (Here we use the notion ``Birkhoff--type
theorem'' for any theorem stating that under certain
 circumstances the gravitational vacuum 
solution has more symmetries than the inserted metric ansatz.) 
The proof is done in two steps: First, the $D$--dimensional
Einstein equation will be reduced to two--dimensional
gravity. Second: By applying the fact that the 
 traceless part of the Ricci tensor identically vanishes 
in two dimensions, we always get an additional Killing vector.

\medskip

As a byproduct, we present a new and coordinate--free 
proof of the classical Birkhoff theorem.

\medskip

The paper is organized as follows: In sct. 2 the warped products
are introduced, in sct. 3 the necessary conformal transformation
is explained, and in sct. 4 the new direct proof of the Birkhoff
theorem is given by an explicit definition of the additional 
Killing vector. 
Sct. 5 shows the  detailed  relation to several 
2-dimensional theories, and sct. 6 gives the summary.

\newpage 

\section{Warped products}

Let us consider a $D$-dimensional Riemannian manifold of 
arbitrary signature and metric $ds^2$ of the form
\begin{equation}
ds^2 \ = \ g_{\alpha\beta}dx^{\alpha}dx^{\beta}  
\ = \ d\sigma^2 \ + \ e^{2U} d\hat\Omega^2
\end{equation}
where
\begin{equation}
d\sigma^2 \ = \ g_{ij} dx^i dx^j 
\end{equation}
is a two-dimensional manifold and 
\begin{equation}
d\hat\Omega^2  \ = \ \hat g_{AB} dx^A dx^B 
\end{equation}
is $n$--dimensional with $n \ge 1$. Hence, $D=n+2 \ge 3$.
The indices $i, \, j$ take the values 0, 1; the indices $A$, 
$B$ take the values 2, \dots , $n+1$; and the indices $\alpha$,
$\beta$
cover both of them, i.e., values 0, \dots, $n+1$.

\medskip

We assume that $\hat g_{AB}$ depends on the $x^A$ only, and
both $U$ and $g_{ij}$ depend on the $x^i$ only. So we have
defined $ds^2$ to be the warped product between 
 $d\sigma^2$ and  $d\hat\Omega^2$ with warping function
$e^{2U}$. {\bf The purpose of the whole consideration is to
show the following: If $ds^2$ is a $D$-dimensional 
Einstein space and the dimension of the isometry group of 
 $d\hat\Omega^2$ equals $k$, then $ds^2$ 
 possesses a $k+1$--dimensional isometry group, at least 
locally.} Taking $d\hat\Omega^2$ as standard two--sphere,
we recover the original Birkhoff theorem. 

\medskip

The method we want to apply for proving this 
is first to reformulate the 
$D$-dimensional Einstein equation for $ds^2$  as a 
field equation in the 2-dimensional space  
 $d\sigma^2$, and second to apply known results [8, 9] 
from scale--invariant gravity in 2 dimensions.  

\bigskip

\section{A conformal transformation}
\setcounter{equation}{0}

To simplify the calculation of the Ricci tensor of
$d s^2$, we perform a conformal transformation 
in $D$ dimensions as
follows: 
\begin{equation}
d \hat s^2 \ = \ \hat g_{\alpha\beta}dx^{\alpha}dx^{\beta} 
 \ = \ e^{-2U} \  d  s^2  
\end{equation}
Using eq. (2.1) this implies 
\begin{equation}
d \hat s^2 
\ = \ d \hat \sigma^2 \ + \  d\hat\Omega^2
\end{equation}
where the analogous conformal transformation in 2
dimensions reads
\begin{equation}
d \hat \sigma^2 \ = \ e^{-2U} \  d  \sigma^2  
\end{equation}
By construction, 
\begin{equation}
  g_{\alpha\beta}
 \ = \ e^{2U} \    \hat g_{\alpha\beta}
\end{equation}
and 
\begin{equation}
 \hat g_{iA} = 0 \qquad \hat g_{ij,A} = 0 
\qquad \hat g_{AB,i} = 0 
\end{equation}
We denote the $D$--dimensional Ricci tensor for $ds^2$
by ${}^{(D)}R_{\alpha\beta}$, and the 
2--dimensional Ricci tensor for $d\sigma^2$ by
$R_{ij}$. For the hatted quantities, we need not
the distinguishing presuperscript because by eqs. (3.2), (3.5), 
the Ricci tensor $\hat R_{ij}$ for $d \hat \sigma ^2$ and the 
Ricci tensor $\hat R_{AB}$ for $d\hat \Omega ^2$
 together form the 
Ricci tensor $\hat R_{\alpha \beta}$ for $d\hat s^2$.
The latter fulfils  $\hat R_{Ai} = 0$ because 
  $d\hat s^2$ represents a direct product. 

\medskip

The Ricci scalars are denoted as follows: 
${}^{(D)}R$ for $ds^2$; $R$ for $d\sigma^2$; 
${}^{(D)} \hat R$ for $d \hat s^2$;
$\hat R$ for $d\hat \sigma^2$; and
${}^{(n)} \hat R$ for $d \hat \Omega^2$.
Applying that $d\hat s^2$ is a direct product
 we get
\begin{equation}
{}^{(D)} \hat R \ = \ \hat R \ + \ {}^{(n)} \hat R
\end{equation}
The assumption that $ds^2$ is an Einstein space reads
\begin{equation}
{}^{(D)}R_{\alpha\beta} \ = \ \Lambda \ g_{\alpha \beta}
\end{equation}
where $\Lambda = \frac{1}{D} \ {}^{(D)}  R$ and 
has a constant value because of $D \ge 3$. 

\medskip

$\Box$ denotes the 2-dimensional D'Alembertian in
 $d\sigma^2$, and $\hat{} \, \Box$ the analogous operator 
 in  $d\hat \sigma^2$. Eqs. (2.2) and (3.3) imply that 
\begin{equation}
\hat g_{ij} \ = \ e^{-2U} \ g_{ij}
\end{equation} 
is the metric for $ d \hat \sigma ^2$. 
Then for the Ricci tensors the
following relation holds: 
\begin{equation}
\hat R_{ij} \ = \ R_{ij} \ + \ g_{ij} \, \Box U
\end{equation} 
i.e.
\begin{equation}
\hat R \ = \ ( R \ + \ 2 \Box U) \, e^{2U}
\end{equation} 
The conformal invariance of the 2-dimensional D'Alembertian
implies
\begin{equation}
\hat{} \, \Box U \ = \ e^{2U} \ \Box U
\end{equation} 
therefore, eq. (3.10) can be rewritten as 
\begin{equation}
 R \ = \ ( \hat R \ - \ 2 \ \hat{} \, \Box U) \, e^{-2U}
\end{equation} 

\bigskip

Now, we turn to the $D$-dimensional conformal transformation
eq. (3.4), we get
\begin{equation}
{}^{(D)}R_{\alpha\beta}  =  \hat R_{\alpha\beta} 
 -  n [U_{ \hat 
; \alpha \beta} - U_{; \alpha} U_{; \beta} ]
- \hat g_{\alpha \beta} [ \  \hat{} \, \Box U + n U_{;\gamma}
U^{ \hat ;\gamma} ]
\end{equation} 
where the ``$\hat ;$''  denotes the covariant derivative 
made with $\hat g_{\alpha \beta}$ and
$$U^{\, \hat ; \, \gamma} \ = \ 
 \hat g^{\, \gamma \,  \mu} \ U_{; \, \mu}$$
First we insert eq. (3.7) into  eq. (3.13) and consider the
$AB$-part of it only. We get after some algebra
\begin{equation}
\hat R_{AB} \ = \ \hat \Lambda \ \hat g_{AB}
\end{equation} 
where we have applied the abbreviation 
\begin{equation}
 \hat \Lambda \ = \ \Lambda e^{2U} \ + \
\hat{} \, \Box U \ + \ n \, U_{;\gamma} \, 
U^{ \hat ;\gamma}
\end{equation}
By construction,  $\hat R_{AB}$ and $\hat g_{AB}$
depend on the $x^A$ only, whereas $\hat \Lambda$
depends on the $x^i$ only. Consequently, 
 $\hat \Lambda$ must be a constant.  
Remark: For $n \ge 3$, this statement follows already from 
the structure of eq. (3.14). For $n = 2$, however, 
it is not obvious from the beginning. 
Eq. (3.14) expresses the fact that $d\hat\Omega^2$
must be a $n$--dimensional Einstein space. 

\medskip

Transvecting eq. (3.14) with $\hat g^{AB}$ we get
\begin{equation}
{}^{(n)} \, \hat R  \ = \  n \,  \hat \Lambda
\end{equation} 

\bigskip

In the next step, we consider the $i,  j$--part of 
eq. (3.13). We get
\begin{equation}
\Lambda g_{ij}  =  \hat R_{ij} 
 -  n [U_{ \hat ; ij} - U_{; i} U_{; j} ]
- \hat g_{ij} [ \  \hat{} \, \Box U + n U_{;m}
U^{ \hat ;m} ]
\end{equation} 
Transvecting eq. (3.17) with $\hat g^{ij}$ we get
\begin{equation}
2 \, \Lambda  \, e^{2U} \ =  \
 \hat R \ - \ D \ \hat{} \, \Box \, U
\ - \ n \, U_{;i} \, U^{ \hat ;i}
\end{equation} 
We apply eqs. (3.10), (3.11) and use the notation 
$$U^{\, , \, i} \ = \ 
  g^{\, i \,  j} \ U_{, \, j}$$
Then we get
\begin{equation}
R \ = \ 2 \, \Lambda 
 \ + \ n \, ( \, \Box \, U \ + \ U_{,i} \, 
U^{,i} \, ) 
\end{equation} 
The remaining of eq. (3.17) consists in the requirement that 
the traceless part of the tensor 
\begin{equation} 
 \hat H_{ij} \ = \  U_{ \hat ; ij} - U_{; i} U_{; j}
\end{equation} 
 has to vanish. 

\bigskip

\section{The 2-dimensional picture (direct way)}
\setcounter{equation}{0}

In this section, we consider gravity in two dimensions
in the two conformally related frames $d\sigma^2$ and
$d\hat \sigma^2$, and the additional scalar field $U$.
The essential equations are (3.15), (3.18) and (3.20):
They represent the field equations in the hatted frame. 
To get the equations  in the unhatted frame we 
use eqs. (3.3) and (3.8). As a result, we have 
 to replace
(3.18) by (3.19), (3.15) by the equation 
\begin{equation}
 \Lambda \ = \ \hat \Lambda e^{-2U} \ - \
 \Box U \ - \ n \, U_{, i} \, 
U^{,i}
\end{equation}
The tracelessness of the term (3.20) will be transformed
to the condition that 
\begin{equation}
 H_{ij} \ = \  U_{  ; ij} + U_{; i} U_{; j}
\end{equation} 
is traceless. One can rewrite eq. (4.2) as
$$
 H_{ij} \ = \  e^{-U} \ ( \,  e^U \, )_{  ; ij}
$$
so that the condition is equivalent to say that
the traceless part of 
$( \,  e^U \, )_{  ; ij}$ must vanish.

\bigskip

At this point we are able to apply appendix B of ref. [8],
where $e^U$ now plays the role of G eq. (B2) of [8].  

\bigskip

First case\footnote{i.e., that case where Schwarzschild
coordinates cannot be introduced for a spherically symmetric
metric}
: Let $U$ be constant in any region of space. 
By eq. (4.1) this case appears only if $\Lambda$ and 
$ \hat \Lambda$ have the same sign. Eq. (3.19) then implies
$R=2\Lambda$. Consequently, $d \sigma ^2$ represents a 
2--surface of constant curvature, and the warped product 
eq. (2.1) specializes to a direct product. 
So we get for this first case:  If 
the dimension of the isometry group of 
 $d\hat\Omega^2$ equals $k$, then $ds^2$ 
 possesses a $k+3$--dimensional isometry group, at least 
locally.

Second case: Let $U_{\, ; \, i } \ne 0$ in any region 
of space. We choose an orientable subregion $W$ of it and fix 
an orientation there. Then we are allowed to use the
 covariantly constant 
antisymmetric pseudotensor $ \epsilon_{ij} $ within $W$. 
It is completely defined by fixing the component
$$
 \epsilon_{01} = \sqrt{\vert \det g_{ij}   \vert} 
$$
Indices will be shifted with the metric 
$ g_{ij} $, and then we define
\begin{equation}
\xi_i \ = \ \epsilon_{ij} \ ( \,  e^U  \, )^{;j} 
\end{equation} 
which is a nowhere vanishing vector field in the region $W$.
 [In index--free notation this can be written as
 \   curl($e^U$) which is analogous to the rot-operator in
three dimensions.]  
Let us calculate its covariant derivative. 
$$
\xi_{i;k} \ = \ \epsilon_{ij} \ ( \,  e^U  \, )^{;j}_{;k} 
$$
Because of eq. (4.2) $( \,  e^U  \, )^{;j}_{;k} $
is $c$ times the Kronecker tensor.
Therefore, we get
\begin{equation}
\xi_{i;k} \ = \ \epsilon_{ij} \ c \ \delta^j_k 
\ = \  c \ \epsilon_{ik}
\end{equation} 
The antisymmetry of the $\epsilon$--pseudotensor  proves
$\xi_i$ to be a Killing vector in  $d\sigma^2$.

But, in contrast to the first case, here it is 
not immediately clear that an isometry of
  $d\sigma^2$ represents an isometry of $ds^2$.

\medskip 

Subcase 2.1.: Let $\xi_i$ be a non--vanishing light-like
vector in any region of space. Then by sct. V A of ref. [8] 
$d\sigma^2$ is flat and of signature $(+-)$. It turns out that 
one of its Killing vectors (not   necessarily 
$\xi_i$ itself) 
 represents also a Killing vector 
of $ds^2$.

Subcase 2.2.: Let $\xi_i$ be a non--light--like
vector in any region of space. Then, by construction, 
it is tangential to the lines of constant values $U$.
Therefore $\xi_i$  represents also a Killing vector 
of $ds^2$.
 
\medskip

In both subcases, we get:  If 
the dimension of the isometry group of 
 $d\hat\Omega^2$ equals $k$, then $ds^2$ 
 possesses a $k+1$--dimensional isometry group, at least 
locally.

\medskip

The boundaries of the cases 1 and 2.1., 2.2.,  
can be covered by a continuity argument.

\bigskip

\section{The 2-dimensional picture  
(way \qquad via \ fourth-order gravity)}
\setcounter{equation}{0}

In this section, we consider gravity in two dimensions
in the two conformally related frames $d\sigma^2$ and
$d\tilde \sigma^2$, and the additional scalar field $U$.
The three essential equations are (3.19), (4.1) and (4.2).
We repeat them here for convenience.
\begin{equation}
R \ = \ 2 \, \Lambda 
 \ + \ n \, ( \, \Box \, U \ + \ U_{,i} \, 
U^{,i} \, ) 
\end{equation}
 
\begin{equation}
 \Lambda \ = \ \hat \Lambda e^{-2U} \ - \
 \Box U \ - \ n \, U_{, i} \, 
U^{,i}
\end{equation} 

and the traceless part of 
\begin{equation}
( e^U )_{;ij}
\end{equation} 
has to vanish. The idea is  to  write the $D$-dimensional
Einstein--Hilbert action for the metric ansatz eq. (2.1).
Eqs. (3.6, 3.12, 3.16) can be subsumed to 
\begin{equation}
{}^{(D)} \,  R = R + n \hat \Lambda e^{-2U} - 
n(n+1)U_{,i} U^{,i} - 2n \Box U
\end{equation} 
The 2--dimensional Lagrangian is obtained after integrating 
the $n$-dimensio\-nal part; the gradient of $U$ can be eliminated
from (5.4) by applying the fact that
$$
(e^{nU} \, U^{,i} )_{;i} = e^{nU} ( \Box U + n U^{,i} U_{,i})
$$
represents a divergence. Then we 
 get the density
\begin{equation}
{\cal L} = [R - (n-1)\Box U - n \Lambda + n \hat \Lambda e^{-2U}
 ] e^{nU} \sqrt{\vert g \vert} 
\end{equation} 
where $g=\det g_{ij}$ is the two-dimensional determinant, and  
$ e^{nU} $ is left over from the $n$-dimensional integration.  
For $n=1$ we can take this form as it stands, but for $n > 1$ 
we apply a conformal transformation
\begin{equation}
\tilde g_{ij} = e^{(n-1)U} g_{ij}
\end{equation} 
 to remove the term with $\Box U$. Analogously to eqs. (3.8), 
(3.12) we get 
\begin{equation}
\tilde R = [R - (n-1)\Box U ] e^{-(n-1)U}
\end{equation} 
and the corresponding Lagrangian scalar 
\begin{equation}
\tilde L = e^{nU} \tilde R - n \Lambda e^U + n \hat \Lambda e^{-
U}
\end{equation} 
fulfils ${\cal L} = \tilde L  \sqrt{\vert \tilde g \vert}$, 
and it holds: Variation of this ${\cal L}$ with respect to $U$
and $\tilde g_{ij}$ gives a system of equations equivalent to 
eqs. (5.1, 5.2, 5.3). In the next step (subsection 5.2.), we 
 try to write this as one single Lagrangian depending on
the two--dimensional metric only, i.e., we want to eliminate
the scalar $U$, and this is possible if one takes a
Lagrangian non-linear in $\tilde R$ as 
will be shown in subsection 5.1.

\bigskip

\subsection{Transformation between fourth and second order}

Here we give a more complete description of that 
 transformation which was sketched in (2.4), (2.5) of ref. [9]
already.  Let $L=L(R)$ be a non--linear Lagrangian in two
dimensions, i.e., $G=\frac{dL}{dR} \ne 0$ (without loss of
generality we assume $G > 0$) and $\frac{dG}{dR} \ne  0$. 
The fourth--order field equation 
following from the variation of $L \sqrt{\vert g \vert}$
with respect to $g_{ij}$ 
 has the trace  
\begin{equation}
0 = GR - L + \Box G
\end{equation} 
and the trace--free part of $G_{;ij}$ must vanish. 
(For dimension 3 or higher, one must instead ensure
the vanishing of the traceless part of 
$G_{;ij} - G R_{ij}$ which is not an equivalent requirement, 
see e.g. ref. [17]. Therefore, the key of the proof 
that eq. (4.3) represents a Killing vector 
is seen to rest on the dimension 2 where the traceless 
part of $R_{ij}$ automatically vanishes.)

\bigskip

Now, we introduce a scalar field $\varphi $ by $G =e^{-2\varphi
}$.
We invert this relation to $R=R(\varphi )$ which is (at least
locally) possible because of our assumptions. Then we define
\begin{equation}
V(\varphi ) \ = \ e^{-2\varphi }R(\varphi ) \ - \
L(R(\varphi )) 
\end{equation} 
and the Lagrangian can now be written as
\begin{equation}
L(\varphi , \, R) \ = \ e^{-2\varphi } \, R \ - \ V(\varphi )
\end{equation} 
If we take this $L(\varphi , \, R)$ as new starting point, then 
the field equations become equivalent:
 variation with  respect to $\varphi $ simply reads
\begin{equation}
0 \ = \ 2 \, e^{-2\varphi }R \ + \ \frac{dV}{d\varphi } 
\end{equation} 
The variation of $L\sqrt{ \vert g \vert}$ with respect to 
the metric gives the trace
\begin{equation}
0 \ = \ V(\varphi ) \ + \ \Box(e^{-2\varphi })
\end{equation} 
and the traceless part of $(e^{-2\varphi })_{;jk}$
vanishes. The two second--order 
eqs. (5.12, 5.13) on the one hand
and the single fourth--order eq. 
(5.9) on the other hand are equivalent. [Let us add the
following: We may overcome the singular point $G=0$ 
by observing that $L$ and $L+\alpha R$ for a constant $\alpha$
give rise to the same fourth--order field equation; the
related $G$ will be increased by $\alpha$, and the field
$\varphi$ changes in a non--linear manner.]

\bigskip

Now we go the other direction: Let $V(\varphi )$ be given
and we take the Lagrangian (5.11). By eq. (5.12) we calculate
$R=R(\varphi )$. First case: This $R$ is a constant.
This takes place if $L= e^{-2\varphi }(R - \Lambda)$ with any
constant $\Lambda$. This is the case where the transformation
to fourth--order gravity becomes impossible.

Second case: This $R$ is not a constant function. This takes
place if the potential $V$ fulfils the inequality
\begin{equation}
\frac{d^2V}{d\varphi ^2} \ + \ 2 \frac{dV}{d\varphi } \ \ne \ 0
\end{equation} 
Then we can invert to $\varphi  = \varphi (R)$ and insert this
into $L(R) = L( \varphi (R),R)$. This is the desired non--linear
Lagrangian leading to a fourth-order field equation equivalent 
that one following from eq. (5.11).

\bigskip

\subsection{Application of this transformation}
 
The transformation given in subsection 5.1. shall now be applied
to the system eq. (5.8) in the tilted version.  Eqs. (5.8) and
(5.11) get the same structure if we put $U=-2\varphi /n$ and
\begin{equation}
V(\varphi ) \ = \ n \Lambda e^{-2\varphi /n} \ - 
 \ n \hat \Lambda e^{2\varphi /n}
\end{equation}
Then the tilted version of eq. (5.12) leads to
\begin{equation}
\tilde R (\varphi ) \ = \ e^{2\varphi } [\Lambda e^{-2\varphi /n}
+ \hat \Lambda e^{2\varphi /n}] 
\end{equation} 
First case: This value $\tilde R$ is constant. (For $n=1$, we
have always $\hat \Lambda =0$, so that $n=1$ is always subsumed 
under this first case.)  Then the transformation to
fourth-order gravity is impossible. 

\medskip

Second case: $n \ge 2$ and either $\Lambda$ or 
$\hat \Lambda$  is non--vanishing. Then inequality (5.14) is
fulfilled and the fourth--order Lagrangian can be given,
but not always in closed form. Therefore we restrict to
two typical examples: 

First example:  $\Lambda = 0$ and $\hat \Lambda = 1$.
We get the Lagrangian
\begin{equation}
\tilde L \ = \ (n+1) \,\tilde R^{1/(n+1)}
\end{equation} 
In such a  context, this Lagrangian was first 
deduced by 
 Rainer and Zhuk, ref. [18, eq. (3.9)].

Second example:  $\Lambda = 1$ and $\hat \Lambda = 0$.
We get
\begin{equation}
\tilde L \ = \ (1-n) \,\tilde R^{1/(1-n)}
\end{equation} 
i.e., in both cases we get scale-invariant fourth-order gravity
where the Birk\-hoff theorem and the corresponding solutions are
 known.

\bigskip

\section{Summary}
\setcounter{equation}{0}

We discussed the warped product 
 metric ansatz (cf. eqs. (2.1), (5.6))
\begin{equation}
ds^2   
\ = \ e^{(1-n)U(x^i)} d\tilde \sigma^2 \ + \ e^{2U(x^i)}
d\hat\Omega^2
\end{equation} 
where 
$$
d\tilde \sigma^2 = \tilde g_{ij}(x^k) dx^i dx^j   \ \ \qquad
i,j,k=0,1 
$$
is a two--dimensional metric and
$$
d\hat \Omega^2 = \hat g_{AB}(x^C) dx^A dx^B \  \qquad A,B,C=
2, \dots n+1
$$
is $n$-dimensional, $n\ge 1$. The question was: Under which
circumstances eq. (6.1) represents an Einstein space in $D=n+2$
dimensions ? First, $d\hat\Omega^2$ has to be an Einstein 
space in $n$ dimensions, i.e., eq. (3.14) has to be fulfilled
with $\hat \Lambda = const. $ (it must vanish for $n=1$).
Second, we introduced the scalar field $\varphi$ via eq. (5.15)
\begin{equation}
U = - 2 \varphi/n
\end{equation} 
and got an equation of 2-dimensional dilaton gravity. 
The latter can be rewritten as fourth--order gravity in two
dimensions. 

\bigskip

To ease reading, we restrict now to the case $\Lambda =0$
and $\hat \Lambda =1$, i.e., $ds^2$ is Ricci-flat, and for 
$d\hat\Omega^2$, metric and Ricci tensor coincide (i.e.
 $n\ge 2$). Then eq. (5.16) implies
\begin{equation}
\tilde R = e^{2\varphi (n+1)/n}
\end{equation} 
Inserting eqs. (6.2), (6.3) into eq. (6.1) we get 
\begin{equation}
ds^2   
\ = \ 
\tilde R^{(n-1)/(n+1)} d\tilde \sigma^2 \ + \
\tilde R^{-2/(n+1)} d\hat\Omega^2
\end{equation} 
Under these circumstances it holds: Metric (6.4) is 
$D$-dimensional Ricci flat if 
$d\tilde \sigma^2$ is a solution  of the 2-dimensional
fourth--order scale--invariant 
gravity following from the Lagrangian (5.17) 
$\tilde L=(n+1)\tilde R^{1/(n+1)}$ 
with $\tilde R \ne 0$. (If $\tilde R < 0$, we have to write 
$\vert \tilde R \vert$ instead of $\tilde R$.)  
And the solutions  of  2-dimensional
fourth--order scale--invariant 
gravity are known in closed form. 
So, we have not only proven a generalized Birkhoff theorem 
for such a warped product  of manifolds, but we have also 
given a procedure how the solutions can be found in closed form. 

\bigskip

To show how this procedure works, let us deduce the spherically
symmetric vacuum solution of Einstein's theory in 4 dimensions.
To this end we put $n=2$ into eq. (6.4) and take  
$d\hat\Omega^2  = d\psi^2 + \sin^2 \psi d \Phi^2$
as standard two--sphere. (That every 
 spherically symmetric 4-metric can be written this way 
with metric (2.1) 
is not proven here, but cf. e.g. ref. [6] for this point.) 
Eq. (5.17) reads $\tilde L=3\tilde R^{1/3}$, this
is the case $k=-2/3$ in the notation of ref. [8], hence, by eqs.
(19, 22) of [8] we get
\begin{equation}
d\tilde \sigma^2 = \frac{dw^2}{A(w)} - A(w)dy^2
\end{equation} 
with $A(w) = C + E \sqrt w$ and 
$\tilde R = \frac{E}{4} w^{-3/2}$. We put $E=4$, and with 
eqs. (6.4), (6.5) we get
\begin{equation}
ds^2   
\ = \ 
\frac{1}{\sqrt w} \left[
\frac{dw^2}{C + 4 \sqrt w}
 - 
(C + 4 \sqrt w) dy^2
 \right]
\ + \ 
 w d\hat\Omega^2
\end{equation} 
With $w=r^2$,
 $y=t/2$, 
$C=-8m$ this 
leads to
\begin{equation}
ds^2   
\ = \ 
\frac{dr^2}{1-\frac{2m}{r}}
 \ - \ (1-\frac{2m}{r})dt^2 
\ + \ 
r^2 [d\psi^2 + \sin^2 \psi d \Phi^2]
\end{equation} 
i.e., the correct Schwarzschild solution. (The generalization 
to higher dimension $n$ is straightforwardly done.) 

\medskip 

The transformation presented here is of a similar structure
to that one presented in [19]; however, in [19] the  
two--dimensional case for $d\sigma^2$ was excluded, 
and here we solely restricted to that two--dimensional case.
In this sense, both approaches are disjoint. The
ansatz eq. (4.3) as Killing vector for 2--dimensional models 
has already be applied several times, e.g. in refs. 
[8] and [20]. 

\medskip

It should be mentioned that at no place the signature of
space-time was used, so the Birkhoff theorem  
is shown to be valid in any signature. The direct way in sct.
4 is equivalent to the way via fourth-order gravity 
in sct. 5. For both ways the whole proof was done in a
 fully covariant manner, i.e., no special coordinates 
 had to be introduced. So we also circumvented the discussion
whether the introduction of Schwarzschild coordinates represents
a loss of generality or not.\footnote{By the way, a 
direct product between two 2-spaces of constant curvature 
represents an example of a spherically symmetric solution of 
Einstein's equation with $\Lambda > 0$ which cannot be
written in Schwarzschild coordinates.}
 The key element of the proof follows from
the fact that the traceless part of the Ricci tensor in
two dimensions identically vanishes, i.e., for the scalar $U$,
the traceless part of 
$( \,  e^U \, )_{  ; ij}$ must vanish. As a consequence, 
$\xi_i \ = \ \epsilon_{ij} \ ( \,  e^U  \, )^{;j} $ 
could 
be proven to be a Killing vector.

\medskip 

\noindent 
{\bf Acknowledgement}. 

\noindent 
Valuable comments by  S. Mignemi and M. Rainer 
 and financial support from the 
 Deutsche For\-schungs\-gemein\-schaft DFG 
are gratefully acknowledged.

\medskip

{\Large {\bf References}}

\bigskip

\noindent 
[1] V. Ruban, Abstracts Conf. General Relativity 
 GR 8 Waterloo,  303 (1977);  
K. Bronnikov and M. Kovalchuk,  {\it 
 J. Phys. A: Math. Gen.} {\bf 13}, 187 (1980);  
 K. Bronnikov and V. Melnikov, 
{\it Gen. Relat. Grav.} {\bf 27}, 465 (1995).

\medskip

\noindent 
[2] H. Bondi and W. Rindler, 
{\it Gen. Relat. Grav.} {\bf 29}, 515 (1997).

\medskip

\noindent
[3] E. Schmutzer, ``Relativistische Physik'', 
Teubner-Verlag, Leipzig 1968; H. Goenner, 
{\it Commun. Math. Phys.} {\bf 16}, 34 (1970).

\medskip

\noindent 
[4] G. Birkhoff and R. Langer,
``Relativity and modern physics'', 
Harvard University Press, Boston 1923; J. Jebsen, {\it 
Ark. Mat. Astron. Fys.} 
{\bf 15}, 18 (1921); W. Alexandrov, {\it Ann. d. Phys. (Leipz.)} 
{\bf 72}, 141 (1923). 

\medskip

\noindent
[5] S. Hawking and G. Ellis, ``The large scale structure 
of space--time'', Cambridge University Press 1973.

\medskip

\noindent
[6] B. Schmidt, {\it Zeitschr. f. Naturforsch.}
 {\bf 22a}, 1351 (1967). 

\medskip

\noindent 
[7] C. Misner, K. Thorne and J. Wheeler, ``Gravitation'',
Freeman, San Francisco 1973. 

\medskip

\noindent
[8] H.-J. Schmidt, {\it J. Math. Phys.} {\bf 32}, 1562 (1991).

\medskip

\noindent
[9]  S. Mignemi and H.-J. Schmidt,
{\it Class. Quant. Grav.} {\bf 12}, 849 (1995). 

\medskip

\noindent 
[10] S. Mignemi,
{\it Ann. Phys. (NY)} {\bf 245}, 23 (1996). 

\medskip

\noindent 
[11] J. Creighton and R. Mann, 
{\it Phys. Rev.}  {\bf D 54}, 7476 (1996).

\medskip

\noindent 
[12] R. Balbinot and A. Fabbri,  
{\it Class. Quant. Grav.} {\bf 13}, 2457 (1996). 

\medskip

\noindent 
[13] M. Katanaev, W. Kummer and H. Liebl,
{\it Phys. Rev.} {\bf D 53}, 5609 (1996).

\medskip

\noindent 
[14] T. Kl\"osch and T. Strobl, 
{\it Class. Quant. Grav.} {\bf 13}, 965 (1996).

\medskip

\noindent 
[15] A. Ashtekar, J. Bicak and B. Schmidt,
{\it Phys. Rev.}  {\bf D 55}, 669 (1997).

\medskip

\noindent 
[16] C. Romero, R. Tavakol and R. Zalaletdinov,
 {\it Gen. Relat. Grav.} {\bf 28}, 365 (1996).

\medskip

\noindent 
[17] H.-J. Schmidt, gr-qc/9703002;  
{\it Gen. Relat. Grav.} {\bf 29}, 859 (1997). 

\medskip

\noindent 
[18] M. Rainer and A. Zhuk,
{\it Phys. Rev.} {\bf D 54}, 6186 (1996). 

\medskip

\noindent
[19] H.-J. Schmidt, {\it Int. J. Mod. Phys.}
 {\bf A 5}, 4661 (1990). 

\medskip

\noindent
[20] T. Banks and M. Loughlin, {\it Nucl. Phys.} 
{\bf B 362}, 649 (1991); 
 S. Solodukhin, {\it Phys. Rev.} {\bf D 51}, 591 
(1995).

\end{document}